# Determination of polariton condensates' critical temperature

E. Rozas[1]*, M. D. Martín[1]*, C. Tejedor[2,3], L. Viña[1,3], G. Deligeorgis[4], Z. Hatzopoulos[4], P. G. Savvidis[4,5]

[1]*Depto. de Física de Materiales e Instituto Nicolás Cabrera, Universidad Autónoma de Madrid (UAM), Madrid, 28049, Spain*
[2]*Depto. de Física Teórica de la Materia Condensada e Instituto Nicolás Cabrera, UAM, Madrid, 28049, Spain*
[3]*Instituto de Física de la Materia Condensada, UAM, Madrid, 28049, Spain*
[4]*FORTH-IESL, P.O. Box 1385, Heraklion, 71110, Greece*
[5]*Dept. of Materials Science & Technology, University of Crete, 71003 Heraklion, Greece, ITMO Univ., St. Petersburg, 197101, Russia*

E-mail: elena.rozas@uam.es, dolores.martin@uam.es

Keywords: photoluminescence, microcavities, polaritons, Bose-Einstein condensation

We investigate the thermal robustness of traveling polariton condensates. We create remote condensates that have never been in contact, and study their interference in momentum space, when they travel with the same velocity, by means of time-resolved photoluminescence. We determine the condensed to thermal, uncondensed polariton fraction, which shows a gradual decay with increasing temperature, and obtain the critical temperature for the Bose-Einstein-like condensate (BEC) phase transition. We tentatively compare our experimental findings with theoretical models, developed for atomic condensates, to describe the condensates' coherence fading with temperature.

## 1. Introduction

Exciton-polaritons present multiple and interesting features due to the strong interaction between their excitonic and photonic components. These quasi-particles can be easily created by optical excitation. Moreover, they can be conveniently confined in low-dimensional structures such as semiconductor microcavities.[1] The polaritons' reduced density of states due to their very low effective mass makes possible their condensation in macroscopic coherent states[2] similar to those of atomic BECs. At low temperatures, these polariton condensates consist of a blend of condensed, coherent particles and thermal, non-condensed polaritons.[3] The coherence has been profusely investigated by the observation of interference effects in traveling condensates, overlapping in real space,[4,5] as well as static condensates.[6-8] Moreover, Antón *et al.* demonstrated the existence of remote coherence between two condensates that were traveling with the same speed and spatially separated by observing the presence of interference fringes in momentum space.[9] It is expected that with increasing temperature, both the spatial and temporal coherence vanish at a critical temperature characteristic of a phase transition. However, the temperature dependence of polariton condensates' coherence has not been investigated in detail.

In this paper, we evaluate the critical temperature of polariton condensates. We employ microcavity ridges where condensates travel ballistically long distances. We analyze the interference fringes appearing when they meet either in real- or momentum-space. From this analysis, we obtain the fraction of condensed to uncondensed polariton populations ($f_C$) at different times after the condensates' creation as a function of temperature. Our results can be fitted with models developed for equilibrium atomic condensates. The critical temperatures obtained at similar delay times in real- and momentum-space are in good agreement. When the condensates experience the presence of excitonic reservoirs the critical temperature is significantly reduced.

## 2. Results and discussion

The sample consists of a high-quality GaAs-based microcavity, with Q ~ 16.000, surrounded by two sets of Bragg mirrors and shows a Rabi splitting of 9 meV. A pattern consisting of ridges with dimensions 20 x 30 µm$^2$ have been sculpted on its surface, from which we select a ridge in a region of the sample with zero detuning. More details about the sample can be found in Ref. [10]. It is mounted in a cold-finger cryostat under vacuum conditions. We establish the temperature range to study the coherence, encompassing the range in which the condensate exists, from 10 K to 42.5 K.

The sample is excited with 2-ps-long light pulses from a Ti:Al$_2$O$_3$ laser. We divide the laser into two beams with different paths verifying that the excitation conditions are identical in both of them. They impinge at the sample surface, through a microscope objective, with the same angle and power density, separated by 70 (±1) µm and zero-time delay between them. The parallel component of the polariton condensate momentum ($k_\parallel$) is related to the angle of detection ($\theta$) by $k_\parallel = n\frac{2\pi}{\lambda}\sin\theta$; the angle of excitation is adjusted to create polaritons with zero momentum, $k_\parallel \sim 0$, at t = 0. The threshold power for condensate´s propagation along the ridge for a pulsed laser source is 3.7 kW/cm$^2$. We set the excitation power at 10 kW/cm$^2$ to perform the experiments, generating a blueshift of 0.6 meV. For further details of the excitation conditions see Ref. [9].

The photoluminescence (PL) at low temperature, 10 K, displays a broad band corresponding to the excitonic emission between 1.5480 and 1.5420 eV and several narrow bands between 1.5420 and 1.5398 eV resulting from the confined lower polariton branches (LPBs). Being aware of the fact that under resonant excitation conditions, the polariton condensates' phase would be obtained from the laser beam coherence, the excitation energy is tuned to the bare exciton level at 1.5459 eV. The measurements are filtered at the energy of traveling condensates, 1.5404 eV, with a resolution of 0.45 meV, using a streak camera with 10 ps of time resolution. We collect the real space as well as the Fourier space emission, focusing the angle-resolved PL just by adding an additional lens to the experimental set up in the latter case. Time-resolved PL traces in real- and momentum-space, as those presented in Figure 1, are the result averaging over millions of laser pulses, allowing to study the dynamics using optical techniques.

In these experiments, two laser beams above power threshold, excite the ridge, creating different condensates traveling along the structure. The dynamics of this system in real space is showed in Figure 1(a). The emission map represents the position along the ridge (x) as a function of time up to 100 picoseconds. The condensates are initially created at the position where the two laser beams impinge, denoted as $B_1$ and $B_2$, located at $\pm 35$ µm at t = 0. From each excitation spot, two condensates emerge (i=1,2), propagating in opposite directions: $L_i$ moving towards left and $R_i$ towards right; the subscript i refers to the excitation spot. The four wave packets (WPs) travel with a constant velocity of $\sim$ 1.5 µm/ps. The trajectories of $L_1$ and $R_2$, the outermost WPs, can only be followed until $\sim$ 45 ps since they escape from the detection area afterwards. On the other hand, $R_1$ and $L_2$ travel towards each other and meet in the center of the ridge (x $\sim$ 0), producing interference fringes as an evidence of spatial and temporal coherence. Weak interference fringes are also perceived along the trajectory of the condensates. The disorder in the sample produces backscattered polaritons that interfere with the condensates, originating fringes with larger periodicity since they travel with a slightly shorter wave vector.[11] The WPs $R_1$ and $L_2$ continue their path without deviating, until they reach the positions $B_2$ and $B_1$ at longer times, $\sim$ 55 ps. At these positions, the WPs decrease their speed and halt at the potential barriers created by the excitonic reservoirs that the condensates cannot overcome. Therefore, the distance between them at these positions is smaller than $70 \pm 1$ µm, as observed in Figure 1(a). They bounce against the barriers and reverse their trajectories, so then $R_1(L_2)$ becomes $\ell_1$ ($r_2$).

Figure 1(b) displays the time-resolved PL emission in momentum space. At high excitation densities, the condensates, initially created at k ∼ 0, acquire a maximum constant momentum in just a few picoseconds. Since the condensates propagate in opposite directions, those moving to the left ($L_1$ and $L_2$) present a negative value of k, -1.3 µm$^{-1}$, while in those moving to the right ($R_1$ and $R_2$) k = +1.3 µm$^{-1}$. Up to 40 ps, the PL shows fringes at k = ±1.3 µm$^{-1}$ due to interferences of pairs of condensates traveling with the same momentum. Beyond 40 ps, condensates still propagating along the structure ($L_1$ and $R_2$) as well as those approaching the potential barriers ($L_2$ and $R_1$) can still be distinguished. The outermost WPs ($L_1$ and $R_2$) continue with the same k value up to 100 ps. $L_2$ and $R_1$, suffer a soft reflection, they reduce their speed in the vicinity of the reservoir at t ∼55 ps, thus, a new set of interference fringes appear around at k = 0. From there on, the WPs reverse their trajectory, $R_1(L_2)$ become $\ell_1$ ($r_2$), and accelerate until they acquire again, the same maximum momenta k = ± 1.3 µm$^{-1}$ as those of the outermost WPs. For condensates moving at negative values of the momentum, interference fringes are observable up to ∼100 ps. At this time, the polaritons arrive to the end of their life span, the emission decays considerably impeding the detection of further fringes.

The interference fringes arise from the remote coherence between polariton condensates that are spatially separated. Yet, a high contribution of non-condensed population is also observed in the PL as a background emission. In this work, we analyze the fraction of condensed to uncondensed polariton populations as a function of lattice temperature, which is not that of the condensates, since they are out-of-equilibrium; however, the dependence on the lattice temperature provides an understanding of the thermal robustness of polariton condensates. We study this temperature dependence for three different sets of fringes in three particular time intervals. In momentum space, we evaluate $f_C$ at $t_1$ (ranging from 13 to 39 ps), when condensates propagate with a maximum value of k, and at $t_3$ (from 42 to 57 ps), when $R_1$ and $L_2$ decrease their momentum to k = 0 in the presence of the reservoir. In real space, we define the interval $t_2$, from 22 to 37 ps, when $R_1$ and $L_2$ meet and overlap in the center of the ridge.

As an example, a trace obtained directly from Figure 1(b), time-integrating in $t_1$, is displayed in Figure 2 (i) for 10 K. It originates from both condensed and non-condensed polariton emission. A large contribution of thermal, uncondensed polariton populations is observed as a background intensity underneath the interference fringes. In order to study and analyze the interference pattern and obtain both contributions to the emission, we perform a comprehensive treatment of the data. The three panels in Figure 2 illustrate the different steps to obtain a neat interferogram to determine through a Fourier analysis the period of the fringes and $f_C$. We start by creating a baseline of the time-integrated profile [dashed line in Figure 2(i)]. The resulting interferogram, after subtracting the baseline, is plotted in Figure 2(ii), where now the interference fringes can be clearly seen. Note that, while the baseline plotted in Figure 2 (i) represents the uncondensed part of the polariton population, the interferogram shown in Figure 2(ii) represents the contribution of only condensed polaritons. The Fourier analysis of the obtained pattern, shown in Figure 2 (iii), provides the amplitude and the period of the interference fringes. After selecting the range $1.2<|k|<2.1$ μm$^{-1}$ in Figure 2 (ii), a peak is found, revealing a period of the fringes in momentum space of $\kappa_0=0.089$ μm$^{-1}$. This period is related to the distance between the condensates created at each excitation spot (d), by $\kappa_0 = 2\pi/d$.[9]

To study the thermal robustness of polariton condensates, we vary the temperature from 10 K up to 32.5 K in steps of 2.5 K. As a compilation, Figure 3 shows the final interferograms, obtained as described above, at different temperatures, for the interval $t_1$. A blue arrow is plotted to indicate increasing temperature. The fringes can be easily observed around ±1.7 μm$^{-1}$ for a wide range of temperatures. Concomitantly to the overall decrease of emission intensity with increasing *T*, a strong decay of the fringes' amplitude is observed up to ~ 32.5 K, where the interferences vanish. Above this temperature, the spatial polariton propagation is limited due to the increasingly faster time decay of the condensates with temperature.

The period of the fringes, revealed by the Fourier analysis, remains constant for all temperatures, $\kappa_0 = 0.089 \pm 0.006$ μm$^{-1}$. Since the distance between two condensates moving in the same direction (d) does not vary in the time interval $t_1$, a $T$ independent period is expected. Thus, we obtain a distance between condensates $d = 71 \pm 5$ μm$^{-1}$, in excellent agreement with the separation between the excitation spots $d = 70 \pm 1$ μm$^{-1}$.

Additional information is found when investigating the interference pattern in time windows $t_2$ and $t_3$. At interval $t_3$, we investigate the fringes produced, in momentum space, when wave packets $R_1$ and $L_2$ decrease their speed overlapping in the vicinity of k ~ 0. At this time, the fringes exhibit an increase of the period, $\kappa_0 = 0.106 \pm 0.002$ μm$^{-1}$, corresponding to a distance in real space of $59 \pm 1$ μm. In fact, as can be induced from Figure 1(a), the real space counterpart PL at $t_3$, shows a smaller distance, $60 \pm 1$ μm, between these condensates, due to the presence of the excitonic reservoir.

We focus now our attention on interval $t_2$, when, in real space, the two traveling condensates $R_1$ and $L_2$ meet at x ~ 0 μm. In this case, the period of the fringes (η) is related to the momentum difference between both wave packets by $\eta = 2\pi / |\vec{k_{R_1}} - \vec{k_{L_2}}|$.[9] The evenness of the excitation conditions, assuring identical propagation speeds of the condensates, leads to $|\vec{k_{R_1}}| = |\vec{k_{L_2}}|$ in the full range of time. We obtain from the Fourier analysis a period of $\eta = 1.8 \pm 0.1$ μm, giving a momentum of $|\vec{k_{R_1 L_2}}| = 1.7 \pm 0.1$ μm$^{-1}$. We experimentally obtain the same maximum k value $|\vec{k_{R_1 L_2}}| = 1.70 \pm 0.02$ μm$^{-1}$, as directly observed in Figure 1(b) for $t_2$.

Now, we evaluate the fraction of condensed to uncondensed polariton populations and obtain its temperature dependence. In Figure 4, we collect $f_C$ as a function of the temperature for real and k-space at the three different time intervals. We compute $f_C$ as the ratio between the area underneath the interference fringes and the baseline [Figure 2 (ii)], and, that enclosed by the baseline. As can be observed in Figure 4 (a), $f_C$ decays with increasing temperature indicating a rise of the relative contribution to the emission of thermal, non-condensed polaritons. Thus, a critical temperature, related to a BEC-like transition, can be determined when $f_C$ vanishes. At $t_1$, when the wave packets travel with the same maximum $|k|$, the experimental data reveals the existence of a 10% of condensed polaritons at low temperature, 10 K. We observe a progressively decrease of this fraction until it vanishes, giving a critical temperature of $34 \pm 3$ K. The temperature range considered in this time interval is limited by the increase of to the signal-to-noise ratio; therefore, for the analysis only temperatures up to 30 K have been considered. Figure 4 (b) shows $f_C$ for the first crossing of the condensates $R_1$ and $L_2$, in real space, in which a similar value of the condensed population is obtained at low temperature. Furthermore, the *T* dependence at $t_2$ is similar to that observed in Figure 4 (a). Due to the link between real- and momentum-space and the fact that intervals $t_1$ and $t_2$ lie in approximately the same range of times, we expect and indeed find similar temperature dependence in both cases: for $t_2$ the fit of the data obtains a critical temperature of $33 \pm 2$ K.

When condensates $R_1$ and $L_2$ halt at the potential barrier at longer times, the calculated fraction in momentum space shows a considerable reduction of the condensed population, as seen in Figure 4(c). In this case, we can only analyze temperatures below 17.5 K due to lifetime constraints. Despite the similarity in the temperature dependence with the previous cases, a faster decrease of $f_C$ yields a lower critical temperature, $24 \pm 4$ K. At $t_3$, the wave packets are located close to the excitonic reservoir, contributing to a quicker reduction of $f_C$ due to both, exciton-polariton scattering and the proximity to the end of their life span.[12-14]

So far, the temperature dependence of the condensates' coherence has been studied theoretically only for condensates in equilibrium. Theoretical models concerning this behavior for non-equilibrium condensates are regrettably lacking. In order to obtain additional information from our experimental findings, we compare our results with two different theoretical models for atomic condensates: a mean-field approach for a purely two-dimensional (2D) atom gas [15] and a three-dimensional (3D) cold atom gas confined in a trap.[16] Both obtain a dependence for the condensed particle fraction ($n_C$) as a function of temperature, given by:

$$n_C(T) = n_0 \left[1 - \left(\frac{T}{T_C}\right)^\beta\right] \quad (1)$$

where $T_C$ represents the critical temperature for the BEC-like transition. In the former model, a linear dependence with temperature, $\beta \sim 1$, is found, while in the latter one, $\beta$ is approximately 3. The fit of our experimental results with both models is presented in Figure 4: the dashed line corresponds to the case $\beta = 3$ while the solid line indicates the linear one. The goodness of the fit is similar in both approaches, giving comparable values of $T_C$ for all three time intervals. Although the $\beta = 3$ fit seems to provide a value of $T_C$ better matching the disappearance of the interference fringes observed for $t_1$ in Figure 3, is hard to conclude which of the two presented theories explains more adequately the evolution of $f_C$ with temperature for non-equilibrium polariton condensates.

On the other hand, in the literature, the models assume a condensed fraction of 1 at zero temperature, a value well above the maximum $f_C$ obtained in our experiments, ~ 0.1. In the case of polariton condensates, their out-of-equilibrium character and the large contribution of non-condensed particles are responsible for the lower values of the condensed fraction as compared with equilibrium condensates.

## 3. Conclusion

In this work, we have proposed an experimental method to evaluate the temperature dependence of the fraction of the condensed to uncondensed polariton population in non-equilibrium condensates. We have observed coherence between condensates created at different excitation positions of a microcavity ridge in real space. This coherence, revealed by the presence of interference fringes, both in real- and momentum-space, is preserved during the polariton propagation in the structure. The condensed particle fraction decreases with increasing temperature. Its vanishing gives the critical temperature of the BEC. Similar values for the critical temperature (~35 K) are found at short times both in real- and momentum-space. A lower temperature is obtained at longer times and it is attributed to lifetime and exciton induced decoherence effects. The results are in good agreement with two different theoretical models developed for the temperature dependence of the condensed fraction of atomic condensates. However, our data do not allow to discern which of them is more appropriate to describe the functional dependence of the condensed fraction on temperature.

**Acknowledgements**

E. Rozas acknowledges financial support from a Spanish FPI Scholarship No. BES-2015-074708. This work was partially supported by the Spanish MINECO Grants No. MAT2014-53119-C2-1-R and No. MAT2017-83722-R. P. G. Savvidis acknowledges support from ITMO Fellowship Program and mega Grant No. 14.Y26.31.0015 of the Ministry of Education and Science of Russian Federation.

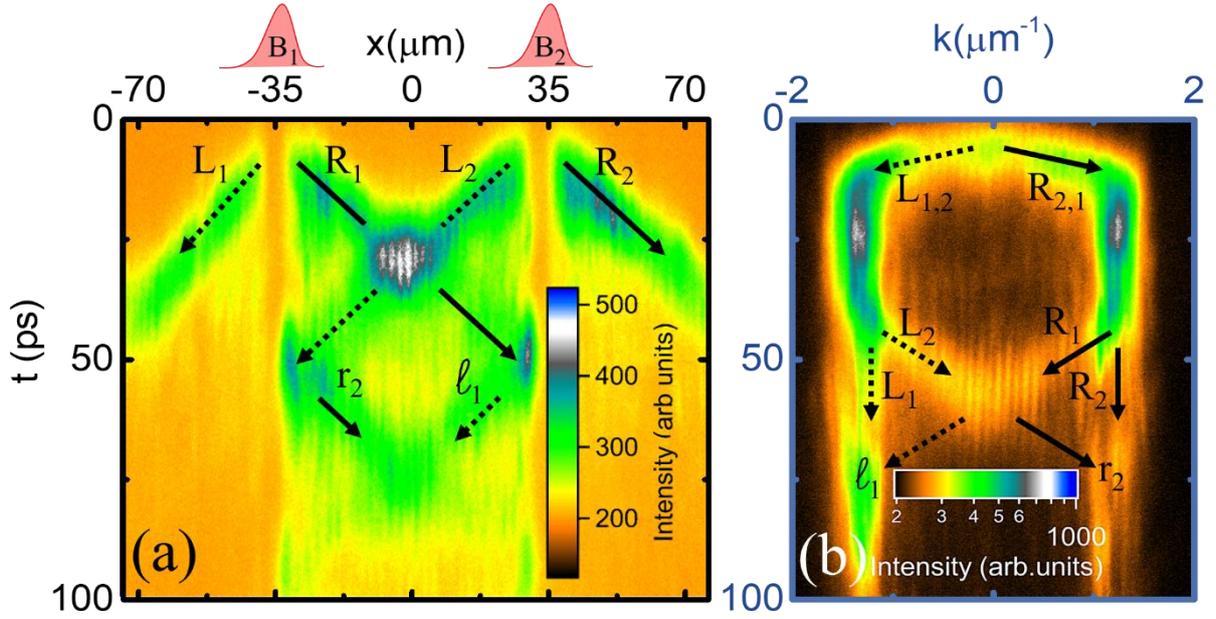

**Figure 1.** (a) Time-resolved PL emission in real space, along the longest dimension (x) of a ridge. The intensity is in a linear false-color scale. The two laser beams impinge on the sample at $B_1$ and $B_2$, separated by 70 μm. The dashed (solid) arrows indicate the WPs moving to the left (right). 1 and 2 refers to the excitation beams. (b) Momentum space emission of (a). The intensity is in a logarithmic false-color scale. Both emissions have been measured at 14 K and with a density power of 6 kW/cm$^2$.

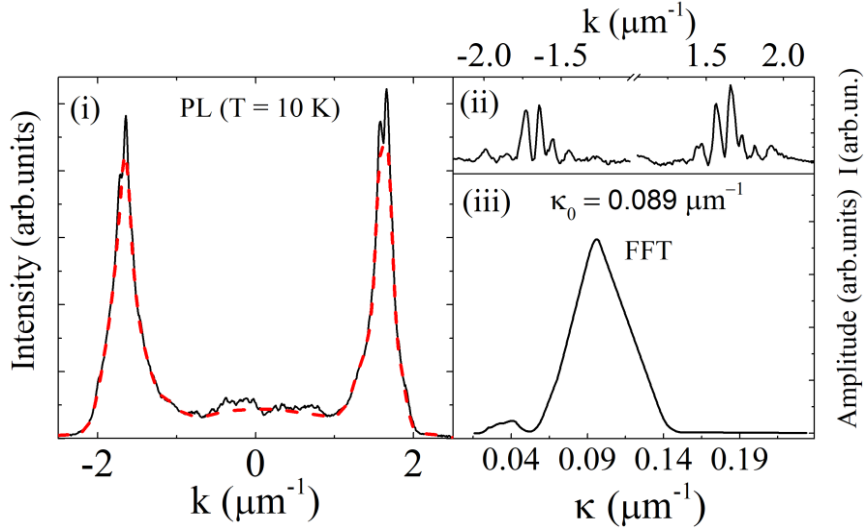

**Figure 2.** (i) Time integrated PL profile versus wavevector k for interval $t_1$ measured at 10 K. The red dashed line depicts the baseline of the emission. (ii) PL profile obtained form (i) exhibiting the interference pattern as a result of the baseline subtraction. (iii) Amplitude of the oscillations shown in (ii), as a function of the period ($\kappa \equiv \Delta k$). The main period of the interference fringes is obtained, $\kappa_0 = 0.089$ μm$^{-1}$.

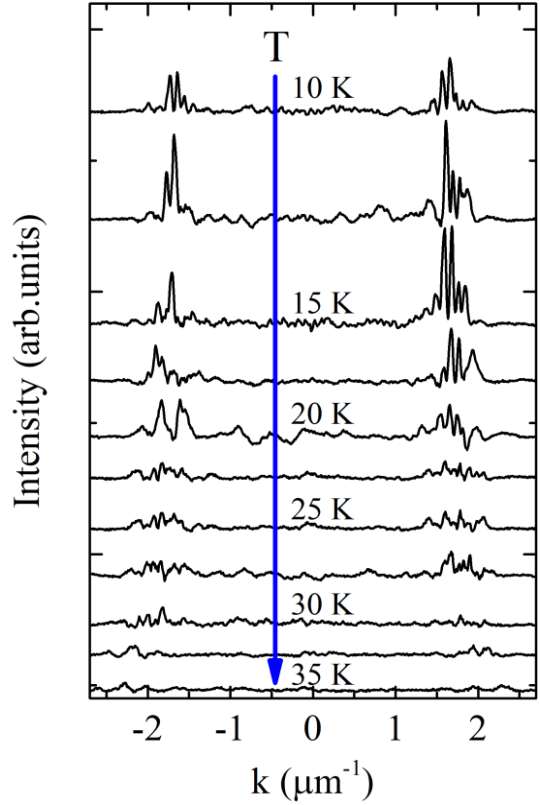

**Figure 3.** Interferogram profile of momentum space at $t_1$ for different temperatures ranging from 10 K up to 35 K, in steps of 2.5 K. The blue arrow indicates the direction of the increasing temperature.

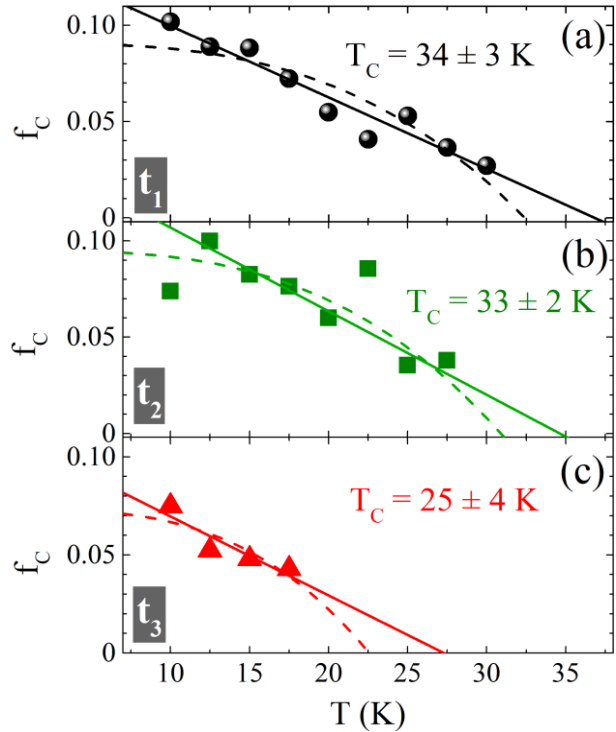

**Figure 4.** Fraction of condensed to uncondensed polariton population as a function of the temperature. (a) and (c) show the calculated fraction in momentum space for intervals $t_1$ and $t_3$, correspondingly. (b) depicts the fraction at the first crossing of the condensates $R_1$ and $L_2$ in real space at $t_2$. Two different fits have been employed: the solid line represents the case for $\beta = 1$ and the dashed line for $\beta = 3$ (see Equation 1).

13